\documentclass{iaus}
\usepackage{graphics}
  \checkfont{eurm10}
  \iffontfound
    \IfFileExists{upmath.sty}
      {\typeout{^^JFound AMS Euler Roman fonts on the system,
                   using the 'upmath' package.^^J}%
       \usepackage{upmath}}
      {\typeout{^^JFound AMS Euler Roman fonts on the system, but you
                   dont seem to have the}%
       \typeout{'upmath' package installed. iaus.cls can take advantage
                 of these fonts,^^Jif you use 'upmath' package.^^J}%
      }
  \else
  \fi

  \checkfont{msam10}
  \iffontfound
    \IfFileExists{amssymb.sty}
      {\typeout{^^JFound AMS Symbol fonts on the system, using the
                'amssymb' package.^^J}%
       \usepackage{amssymb}%
       \let\le=\leqslant  
         
      }{}
  \fi

  \IfFileExists{amsbsy.sty}
    {\typeout{^^JFound the 'amsbsy' package on the system, using it.^^J}%
     \usepackage{amsbsy}}
    {\providecommand\boldsymbol[1]{\mbox{\boldmath $##1$}}}

  \usepackage{psfig}

\newcommand\etal{\mbox{\textit{et al.}}}

\newcommand\eg{e.g.\ }
\newcommand\cf{c.f.\ }

\def\gsim{\mathrel{\raise0.35ex\hbox{$\scriptstyle >$}\kern-0.6em % Greater/squiggles
\lower0.40ex\hbox{{$\scriptstyle \sim$}}}}
\def\lsim{\mathrel{\raise0.35ex\hbox{$\scriptstyle <$}\kern-0.6em % Less than/squiggles
\lower0.40ex\hbox{{$\scriptstyle \sim$}}}}
\def\halpha{{\rm H$\alpha$}}

\def\oii{{\rm O{\sc ii}}}

\title[Outskirts of Galaxy Clusters: intense life in the suburbs]
      {The Mass Assembly History of Galaxies and Clusters of Galaxies}

\author[T. Kodama {\it et al.\/}]%
{Tadayuki Kodama$^1$, the PISCES team \and the SXDS team}

\affiliation{$^1$National Astronomical Observatory of Japan, Mitaka,
Tokyo 181--8588, Japan\\ email: kodama@th.nao.ac.jp}

\pubyear{2004}
\volume{195}
\pagerange{1--7}
\date{?? and in revised form ??}
\jname{Outskirts of Galaxy Clusters: intense life in the suburbs}
\editors{A. Diaferio, ed.}
\begin{document}

\maketitle

\begin{abstract}

We discuss the mass assembly history both on cluster and galaxy scales and
their impact on galaxy evolution.

On cluster scale, we introduce our on-going PISCES project on Subaru,
which plans to target $\sim$15 clusters at $0.4\le z\le 1.3$ using the
unique wide-field (30$'$) optical camera Suprime-Cam and the spectrograph
both in optical (FOCAS, 6$'$) and near-infrared (FMOS, 30$'$).
The main objectives of this project are twofold: (1) Mapping out the large
scale structures in and around the clusters on 10--14~Mpc scale to study the
hierarchical growth of clusters through assembly of surrounding groups.
(2) Investigating the environmental variation of galaxy properties along
the structures to study the origin of the morphology-density and
star formation-density relations. Some initial results are presented.

On galactic scale, we first present the stellar mass growth of cluster
galaxies out to $z\sim1.5$ based on the near-infrared imaging of distant
clusters and show that the mass assembly process of galaxies is largely
completed by $z\sim1.5$ and is faster than the current semi-analytic
models' predictions.
We then focus on the faint end of the luminosity function at $z\sim1$
based on the Subaru/XMM-Newton Deep Survey imaging data.
We show the deficit of red galaxies below M$^*$+2 or 10$^{10}$~M$_{\odot}$,
which suggest less massive galaxies are either genuinely young or still
vigorously forming stars in sharp contrast to the massive galaxies where
mass is assembled and star formation is terminated long time ago.

\end{abstract}

\firstsection % if your document starts with a section,
%\section{Introduction}

%\vspace{-0.2cm}
\section{Panoramic Imaging and
Spectroscopy of Cluster Evolution with Subaru (PISCES)}

\begin{figure}
\centerline{
\psfig{file=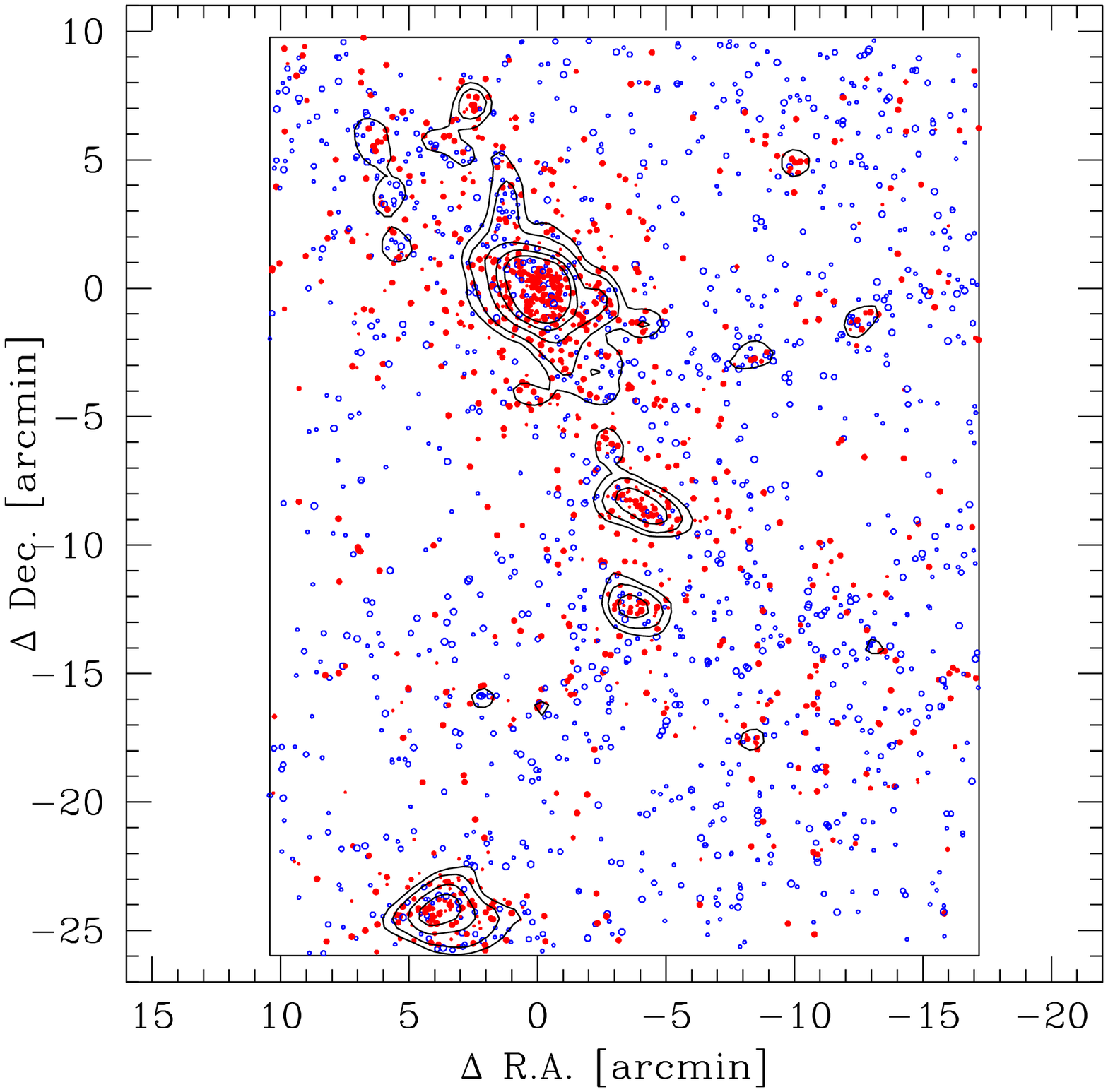,angle=0,width=6.5cm}
\psfig{file=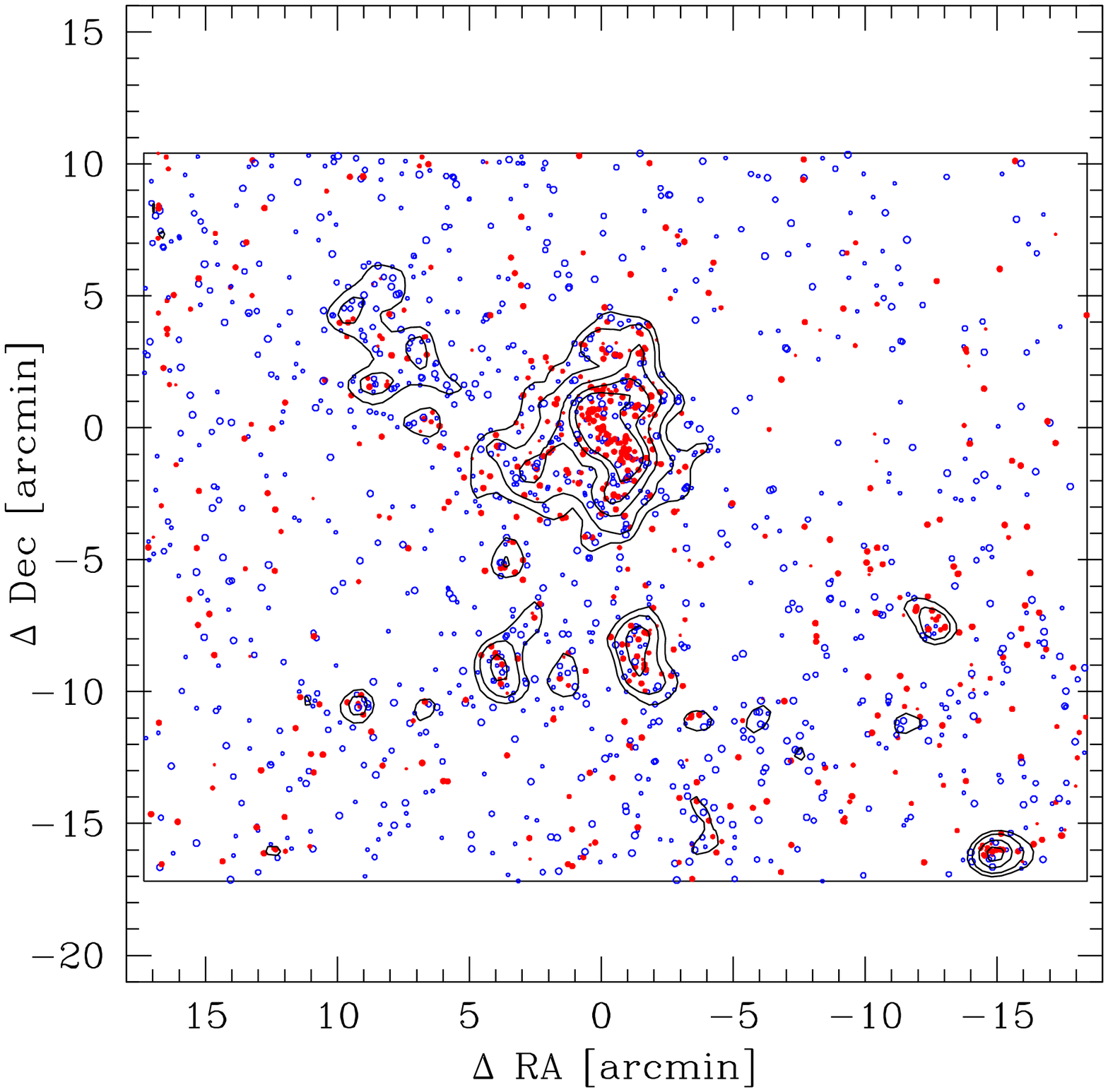,angle=0,width=6.5cm}
}
\caption{
The panoramic maps of
CL0016+16 cluster ($z$=0.55; left panel)
and RXJ0152.7$-$1357 ($z$=0.83; right panel).
10~arcminutes correspond to physical scales of 3.8 and 4.6~Mpc,
respectively.
Using photometric redshift technique based on multi-colour data
($BVRi'z'$ and $VRi'z'$, respectively),
plotted are the photometric member candidates selected with redshift cuts of
0.50$\le$$z$$\le$0.58
and 0.78$\le$$z$$\le$0.86,
respectively.
Contours show local 2-D number density of galaxies at
1.5, 2, 3, 4, 5 $\sigma$ above the mean density.
Coordinates are shown relative to the centre of the main cluster.
Large scale filamentary structures ($>$10Mpc) are seen in both clusters.
}
\label{fig:map_photz}
\end{figure}

\begin{figure}
\centerline{
\psfig{file=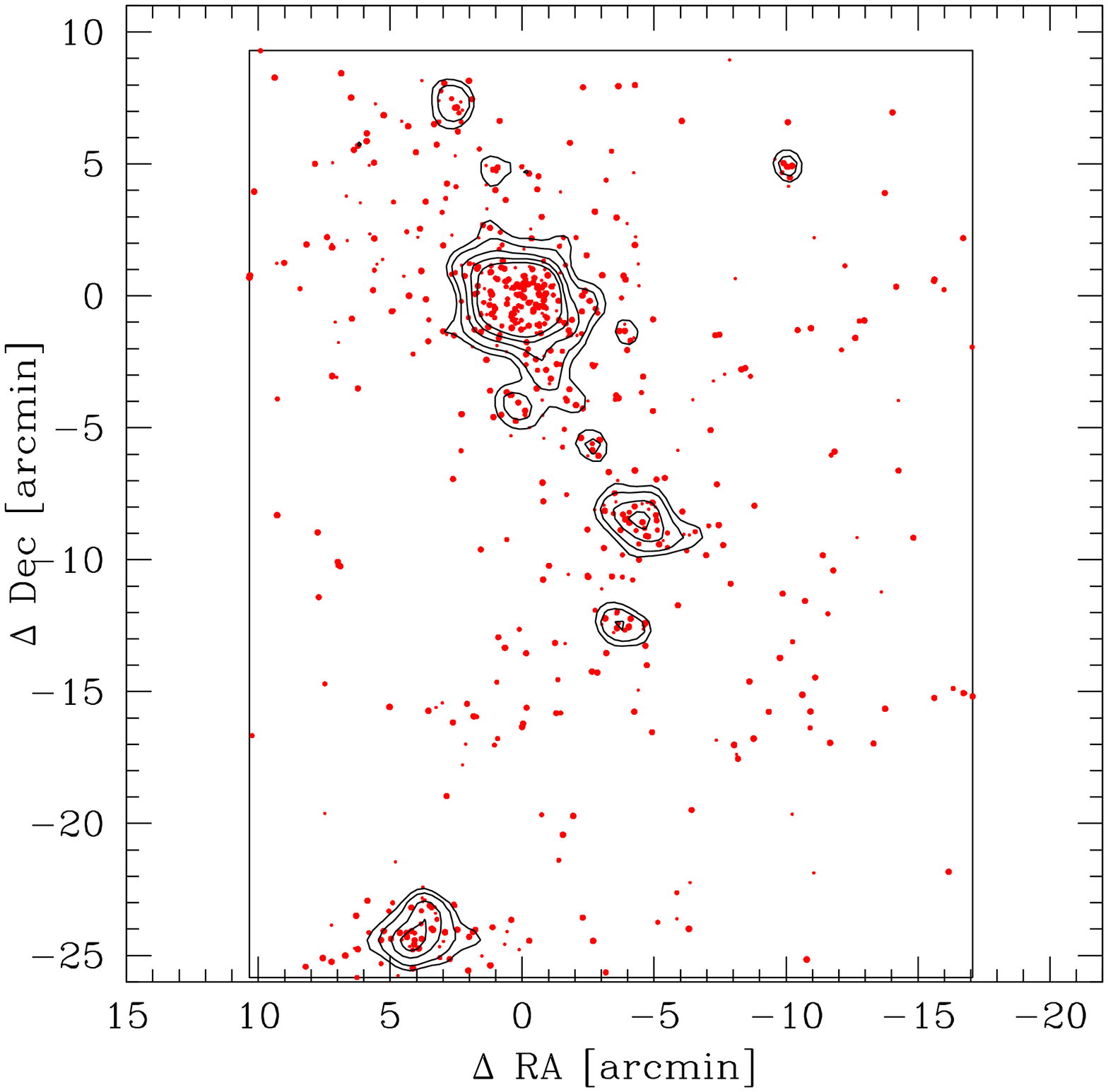,angle=0,width=6.5cm}
\psfig{file=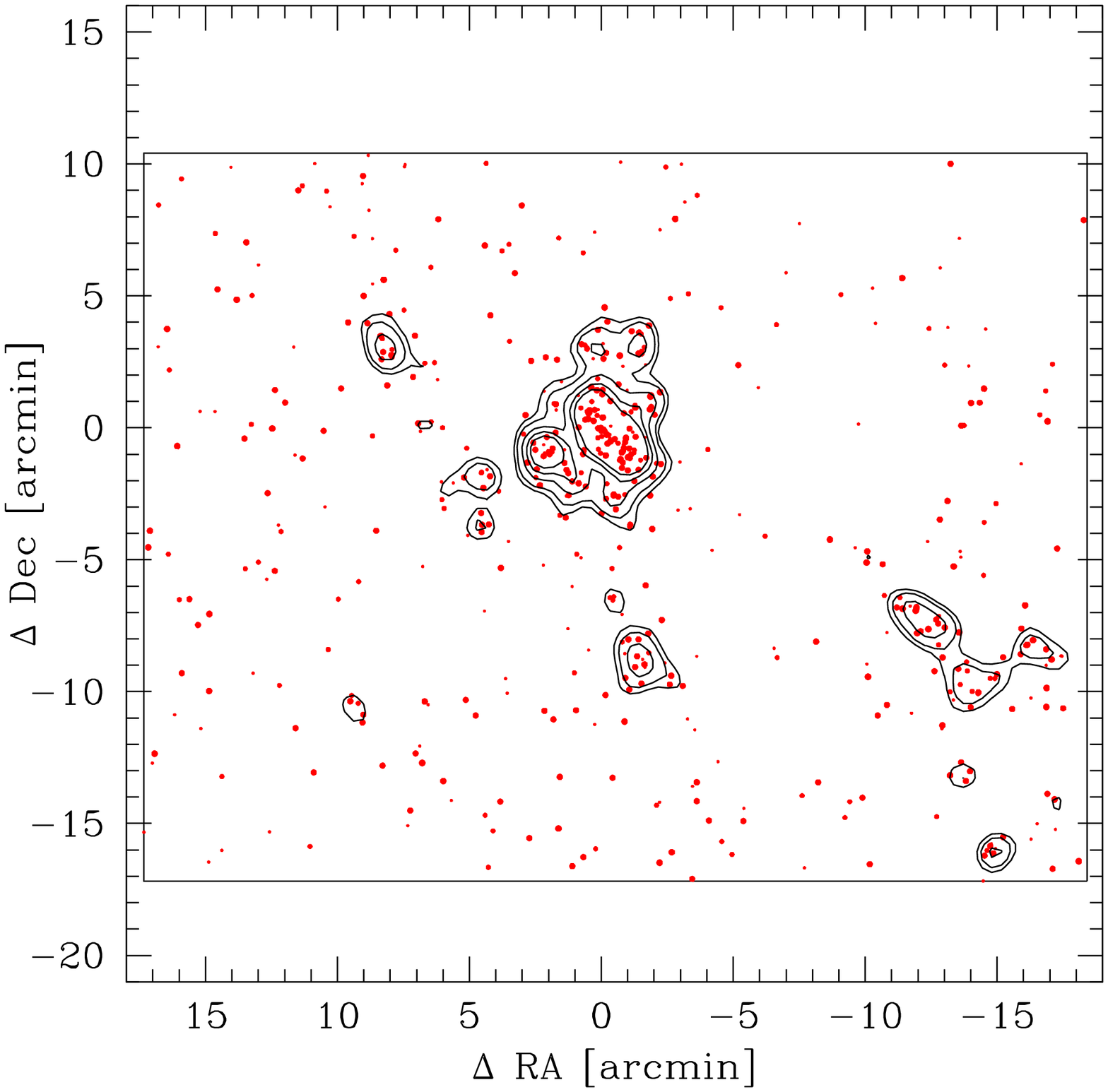,angle=0,width=6.5cm}
}
\caption{
The same as Fig.~\ref{fig:map_photz},
but only red galaxies on the colour-magnitude sequence 
are shown rather than phot-$z$ selected galaxies.
The red galaxies are selected on the basis of
$VRi'$ colours for CL0016+16 and $Ri'z'$ colours
for RXJ0152.7$-$1357, respectively,
and some red colour slice cuts are applied to isolate the passively evolving
galaxies at cluster redshifts.
This technique gives narrower redshift slice ($\Delta$$z$$\sim$0.05) hence
has less projection effect, but tends to be biased to the systems dominated by
red populations. It is therefore complemental to the phot-$z$ slice technique
used in Fig.~\ref{fig:map_photz}.
}
\label{fig:map_red}
\end{figure}

PISCES is a panoramic imaging survey of distant clusters using the
Subaru wide-field optical camera Suprime-Cam which provides
34$'$$\times$27$'$ field of view corresponding to a physical area of
16$\times$13~Mpc$^2$ at $z\sim1$.
This long term project has started since 2003, and
we aim to target $\sim$15 X-ray selected distant clusters in total at
$0.4\lsim z\lsim 1.3$, in good coordination with {\it ACS/HST},
{\it XMM}, and {\it Chandra} observations.
%%The goals of this programme is to map out the large scale structures
%%around clusters and trace the cluster assembly history and then to look into
%%the galaxy properties as a function of environment along the structures
%%to directly identify the environmental effects acting on galaxies
%%during their assembly to higher density regions.
This unique project is currently underway and some preliminary
results on the large scale structures over the entire Suprime-Cam fields
are shown in Figs.~\ref{fig:map_photz} and \ref{fig:map_red}.
These two rich clusters at $z$=0.55 and 0.83 were imaged in multi optical
bands, and photometric redshifts (\cite{k99}) have been applied to
efficiently remove foreground/background contaminations and isolate
the cluster member candidates (\cf \cite{k01}).
Many substructures are then clearly seen around the main body of the
clusters which tend to be aligned in filamentary structures extending to
$>$10~Mpc scale across.
Although these structures should be confirmed spectroscopically later on,
these already provide good evidence for cluster scale assembly in the
hierarchical Universe.

\begin{figure}
\centerline{
\psfig{file=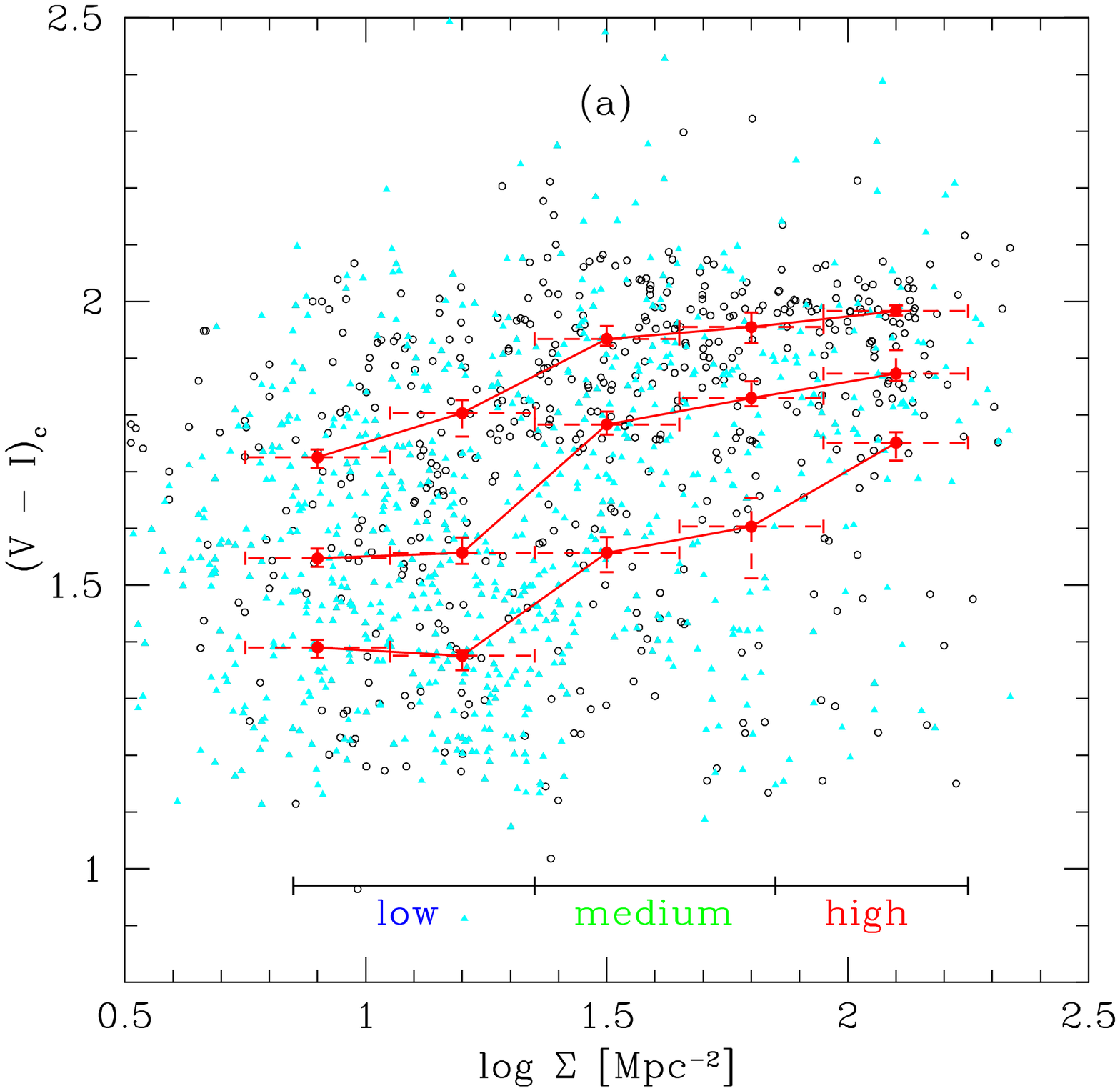,angle=0,width=6.5cm}
}
\caption{
The variation in colour versus local galaxy density,
for phot-$z$ selected cluster members brighter than $I=23.4$
in CL0939+47 cluster at $z=0.41$ (\cite{k01}).
The open circles and filled triangles show the galaxies brighter or fainter
than $I=21.4$ ($M_V^{\ast}$+2), respectively.
The three red lines represent the loci of the 25, 50, and 75th percentile
colours.  The local number density is
calculated from the 10 nearest galaxies and we correct this for residual
field contamination in the photometric members using the blank field data.
}
\label{fig:colden}
\end{figure}

The large scale structures that we see around the clusters provide us
the unique opportunities to look into the environmental effects on galaxies
as they assemble to denser regions.
Kodama \etal\, (2001) have presented the environmental dependence of galaxy
colours along the filamentary structures around the CL0939+47 cluster at
$z$=0.41.  They have shown that the galaxy colours change rather sharply
at relatively low density regions such as galaxy groups along the filaments
well outside of the cluster core where the galaxies have not yet passed the
central region of the clusters yet (see also \cite{gray04}; \cite{treu03}).
Together with the similar findings
in the local Universe (\cite{lewis02}; \cite{gomez03}).
the environmental effects that truncate star formation are not driven by
the cluster specific mechanism such as ram-pressure stripping (\cite{abadi99})
and but are found to be much wider spread into low density regions.
We should therefore pay greater attention to galaxy groups
as the key hierarchy for the environmental effects and try to identify what
is happening on galaxies in this environment (see below).
%(see \S~1.1).
It is also important to extend this analysis to higher redshifts as the
galaxy environment should change dramatically during the course of vigorous
assembly, which is probably related to the appearance of morphology-density
relation (\cite{d80}).

%%\subsection{Follow-up plans for PISCES}

Obviously, taking as many spectra as possible from the photometrically
identified large scale structures is crucial to prove their reality,
since our photometric approach may well suffer from the projection
effect along the line of site as we go to lower density regions
due to the broad phot-$z$ slice cuts that we apply.
Importantly, \oii\, line and/or the 4000\AA\, break feature are detectable
for our PISCES targets in the optical spectroscopy (such as FOCAS) and
\halpha\, line comes to the FMOS window (0.9$\mu$m$<$$\lambda$$<$1.8$\mu$m).
Not only definitively removing the foreground/background contamination
and identifying the physically associated real groups,
spectroscopic redshifts of individual cluster members will also provide us
two critical information: (1) Dynamical mass of the systems which can then
be compared to lensing mass and the X-ray mass to address the dynamical
state of the systems. (2) 3-D velocity structures, providing the recent
and/or near future cluster-cluster/cluster-group merger histories
(\eg \cite{czoske02}).

Also, \oii\, and \halpha\, lines will offer the measures of on-going star
formation rate of galaxies.  (The latter is preferred since it is much less
affected by dust extinction or metallicity variation (\cite{kenn84})).
Therefore we can directly identify when, where and on what timescale
the star formation is truncated as the galaxies/groups fall into clusters
along the filaments (\eg \cite{k04b}).
Moreover, we will combine the information from other spectral indices
such as Balmer lines and colours in order to resolve the
recent star formation histories in galaxies in fine time scales.
Different spectral features are sensitive to different stellar ages
(\cite{kenn84}; \cite{balogh99}; \cite{bia99}):
the emission lines (\oii\, \halpha) will measure the amplitude of on-going 
star formation (10$^7$ yrs), while the Balmer absorption line
give the luminosity (mass) contribution from the stars formed immediately
before the truncation (10$^{8-9}$ yrs) (\cite{d92}; \cite{couch87})
and the 4000\AA\, break and broad-band colours specify the
features of longer-lived populations ($>$10$^9$ yrs) (\cite{kb01}).
Resolving the recent star formation histories in galaxies in the
transition regions (groups) is the key to understanding the
physical processes behind the truncation.
In particular, the existence of strong nebular emissions 
would support the galaxy-galaxy mergers which trigger star-burst,
and very strong Balmer absorption line (E+A or k+a) would follow in the
post star-burst phase (\cite{bia99}).
If the star formation is more gradually truncated ($\gsim$1Gyr) due
to halo gas stripping (strangulation), we would not see any excess of
E+A/k+a features (\cite{balogh99}).

It is also important to investigate the morphologies of the galaxies in
these groups.
The key question here is whether the transformation of morphologies is driven
by the same mechanisms as those responsible for the truncation of star
formation (\cite{bia99}; \cite{treu03}).
Furthermore, stellar mass function of galaxies (see \S2) in groups compared to
other environment will provide information on galaxy-galaxy
merger in this hierarchy since the mergers increase the fraction of massive
galaxies.  On the contrary, strangulation does not change mass, hence can be
distinguished by this test.

\vspace{-0.2cm}
\section{Stellar mass assembly of massive galaxies in high-$z$ clusters}

\begin{figure}
\centerline{
\psfig{file=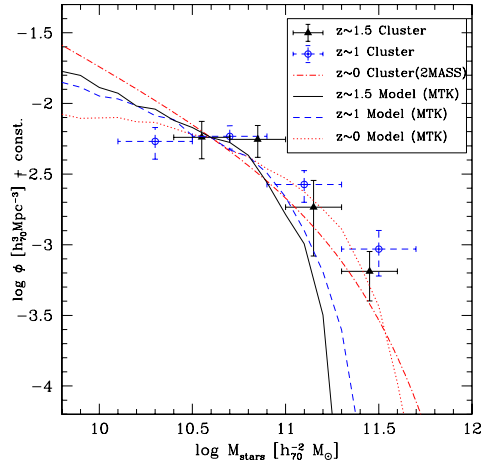,angle=0,width=6.5cm}
}
\caption{
Stellar mass functions of galaxies in clusters as a function of
redshift.
The Kennicutt's initial mass function (\cite{kenn83}) is used to scale the
stellar mass.
The open diamonds and the filled triangles show
the stellar mass functions for $z\sim1$ and $z\sim1.5$ clusters,
respectively, which are compared to the local counterpart from the
2MASS survey (dot-dashed curve).
All the curves and the data points are normalized at
5$\times$10$^{10}$M$_{\odot}$ so as to have the same amplitude.
The theoretical predictions from a semi-analytic model (\cite{n02})
are presented for comparison, which are made for galaxies in the haloes
whose circular velocities are greater than 1000~km/s at each epoch. 
The mass assembly of massive galaxies in the real Universe is much
faster than the hierarchical model prediction.
}
\label{fig:mf}
\end{figure}

We now move on to the galactic scale, and firstly we present the stellar mass
functions of galaxies in high-$z$ clusters constructed from deep
near-infrared imaging in $J$ and $K_s$ (Fig.~\ref{fig:mf})
(\cite{kb03}; \cite{k03b}).
The stellar mass function of galaxies derived from the $K$-band
observations is a good tracer of mass assembly history of galaxies
and therefore provides a critical test for the CDM-based bottom-up picture
(\eg \cite{kc98}; \cite{baugh02}).

We have combined two $z\sim1$ clusters (3C336 and Q1335+28; \cite{kb03})
and five $z\sim1.5$ clusters (Q0835+580,
Q1126+101, Q1258+404, Q0139--273, and Q2025--155; \cite{hall98}; 2001;
\cite{best03}) to increase statistics.  We have subtracted the control
field counts taken from the literature (\cite{s99}; \cite{s01}; \cite{best03}).
Applying the same technique described in Kodama \& Bower \etal\, (2003),
we construct the field-subtracted stellar mass functions
of galaxies in high-$z$ clusters, primarily using $K_s$-band flux and
also using $J-K_s$ colour as a measure of the M/L ratio.
As shown, little evolution is observed since $z=1.5$ to the present-day
(2MASS clusters; \cite{balogh01}),
indicating that the mass assembly on galaxy scale is largely completed
by $z\sim1.5$ in the cluster environment.
This epoch of mass assembly of massive galaxies is earlier than the
prediction of the hierarchical models as shown for comparison (\cite{n02}).

It is interesting to apply similar analyses for the general field, since the
high density regions may be the special places where galaxy formation
processes take place in an accelerated way.
Recently, Pozzetti \etal (2003) and Glazebrook \etal (2004)
investigated the stellar mass growth of massive galaxies out to $z\sim1.5$
based on the large (mostly) spectroscopic sample in {\it K20}
(52 arcmin$^2$) and {\it GDDS} (120 arcmin$^2$), respectively, and found
no or little evolution is seen in the stellar mass of massive galaxies,
suggesting the early assembly of galaxies even in the general field
environment.

It is also important to go even higher redshifts
to firstly identify the epoch of assembly of massive galaxies when
they start to break down into pieces.
Recent deep NIR observations in fact start to enter such formation epoch
(\eg \cite{dick03}; \cite{franx03}; \cite{rud03}; \cite{daddi03}),
and this aspect will be further extended by on-going and near future space
missions such as {\it SIRTF} (\cite{dick02}; \cite{pozz03})
and {\it Astro-F} (Japanese mission).

\vspace{-0.2cm}
\section{Down-sizing in galaxy formation seen at $z\sim1$}

\begin{figure}
\centerline{
\psfig{file=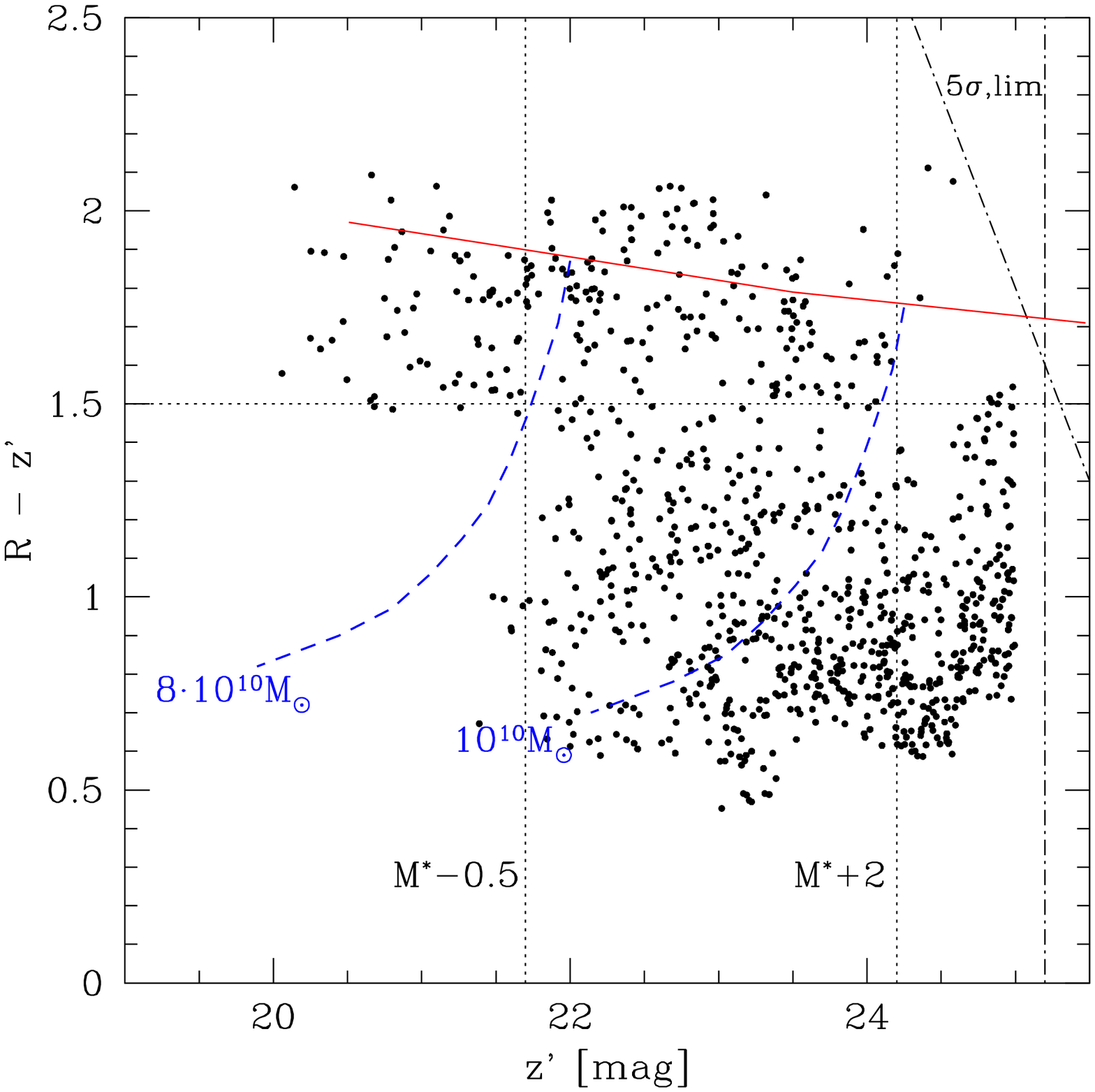,angle=0,width=6.5cm}
}
\caption{
Field-corrected colour-magnitude diagram for the $z\sim1$
galaxies in high density regions in SXDS (\cite{k04}).
The solid line show the expected location of colour-magnitude sequence
at $z\sim1$ assuming a passive evolution with $z_{\rm form}$=5 (\cite{k98}).
The deficit of both blue galaxies at the
bright/massive end and the deficit of red galaxies at the faint/less-massive
end are both clearly identified.
The stellar masses are scaled using the Kennicutt's IMF (\cite{kenn83}).
}
\label{fig:z1cm}
\end{figure}

It is known that the galaxy properties depend on mass or luminosity as well.
Using the SDSS data, Kauffmann \etal\, (2003) and Baldry \etal, (2004)
have shown an interesting break mass at
3$\times$10$^{10}$M$_{\odot}$ above which the dominant population is
red passively evolving galaxies, while below that mass the contribution of
blue active galaxies become dominant.
Morphological mix of galaxies is also known to be strongly luminosity
(or stellar mass) dependent (\eg \cite{treu03}).
It is therefore indicative that the formation of massive galaxies and
less massive galaxies are quite different, in the sense that massive
and/or early-type galaxies form early in the Universe, while dwarf and/or
late-type galaxies form later on average or still forming stars at present.
The mass dependent star formation history is referred to as
``down-sizing'' (\cite{cowie96}).

Given the down-sizing effect in the local Universe, as we look back into
higher redshift Universe, we expect to go beyond the formation epoch of
small galaxies or towards the early stage of their formation,
as well as to approach the formation epoch of more massive galaxies.
To test this hypothesis, we investigate the galaxy colours at $z\sim1$
as a function of luminosity (or stellar mass) utilizing the unique
Suprime-Cam imaging data-set ($BRi'z'$) on the Subaru/XMM-Newton Deep
Survey (SXDS).
These data are both sufficiently deep ($z'_{AB}$=25, 6-10$\sigma$) and
wide (1.2 deg$^2$), which enable us for the first time to investigate
the photometric properties of statistical sample of galaxies at $z\sim1$
down to $\sim$$M^*$+3 with respect to the passive evolution.
We first identify five $z\sim1$ high density regions by applying colour cuts
at 1.7$<$$R$$-$$z'$$<$2.0 and 0.8$<$$i'$$-$$z'$$<$1.1, which correspond
to the colours of passively evolving galaxies at $z\sim1$ (\cite{k98}).
We then combine these five regions 
(amounting to 141 arcmin$^2$ in total) and subtract off the low density
regions at $z\sim1$ sampled from the same data-set and scaled to the same
area.  We do this subtraction on the colour-magnitude diagram (\cite{kb01})
and isolate the $z\sim1$ galaxies in a statistical sense.  This method
should work since both high and low density regions at $z\sim1$ are expected
to have the same amount of foreground/background contaminations.

The field-corrected colour-magnitude diagram thus constructed for $z\sim1$
galaxies in high density regions is shown in Fig.~\ref{fig:z1cm}.
The most striking feature in the galaxy distribution on this diagram is
that the galaxies are separated into two distinct populations,
`bright+red' and `faint+blue'.
More precisely, we show a deficit of red and faint galaxies
below $M^*$+2 or 10$^{10}$M$_{\odot}$ in stellar mass
and a lack of blue massive galaxies
above $M^*$$-$0.5 or 8$\times$10$^{10}$M$_{\odot}$ in stellar mass.
The down-sizing in star formation is therefore also seen at high redshift,
where star formation in massive galaxies takes place early in the Universe
and is already truncated by $z\sim1$, while almost all of small galaxies
are still forming stars at $z\sim1$ (\cite{k04}; \cite{bia04};
see also De Lucia \etal\, in this volume).
Together with the early-assembly of massive galaxies see in \S~2,
galaxy formation does in fact take place in ``down-sizing'' fashion
as apparently opposed to the ``bottom-up'' scenario.
Some critical physical mechanisms in galaxy formation may be still
missing in the current models.

It is important to note that there are two possible interpretations
for the blue colours of faint galaxies.
One is that they are just forming, and the other is that they have much
extended star formation over the long period of time.
To separate out these two possibilities, the key quantity to measure is
the so-called birth-rate parameter ($b$) of these faint blue
galaxies (\cite{kenn88}).
Here $b'$ is re-defined as on-going star formation rate divided by
integrated stellar mass of the system ($b'$=$SFR$/$M_{\rm stars}$).
$SFR$ can be derived from \oii\, and \halpha\, line intensities
by spectroscopy or narrow-band imaging of the PISCES clusters and the SXDS
field while $M_{\rm stars}$ can be obtained from the deep NIR imaging.
This parameter describes what phase of star formation each galaxy is in,
and therefore can discriminate between genuinely young galaxies
formed recently (large $b'$) and star forming galaxies formed long
time ago (small $b'$).
It is also interesting to directly detect the brightening of the break
luminosity or mass with increasing redshift by going even
higher redshifts.

%\vspace{0.3cm}
%We have assumed the cosmological parameters of
%($H_0$, $\Omega_m$, $\Omega_{\Lambda}$)=(70, 0.3, 0.7), throughout this
%paper.
%\begin{acknowledgments}
%This work was financially supported in part by a Grant-in-Aid for the
%Scientific Research (No.\, 15740126) by the Japanese Ministry of Education,
%Culture, Sports and Science.
%\end{acknowledgments}

\end{document}